\documentclass[prd,aps,twocolumn,superscriptaddress,preprintnumbers,notitlepage,nofootinbib]{revtex4-2}
\usepackage{amssymb,amsthm,amsmath}
\usepackage[utf8]{inputenc}
\usepackage{textcomp}
\usepackage{color}      
\usepackage{slashed}    
\usepackage{verbatim}
\usepackage[normalem]{ulem} 
\usepackage{hyperref}
\usepackage{soul}
\usepackage{svg}
\usepackage{cancel}

\usepackage{graphicx}
\usepackage{subfigure}

\usepackage{CJKutf8}

\usepackage{booktabs}
\usepackage{rotating}   
\usepackage{multirow}   
\usepackage{simpler-wick}
 \usepackage[compat=1.1.0]{tikz-feynman}
\usepackage{amsmath}
\usepackage{axodraw2}
\tikzfeynmanset{compat=1.1.0}
\usetikzlibrary{decorations.markings}

\begin{document}
\begin{CJK*}{UTF8}{gbsn}

 \title{Complex $\tau$ Electric Dipole Moment from GeV-Scale New Physics }

	\author { Zhong-Lv Huang(黄钟吕)}\email{huangzhonglv@sjtu.edu.cn}

		\affiliation{
		State Key Laboratory of Dark Matter Physics,
		Tsung-Dao Lee Institute \& School of Physics and 
		Astronomy, Shanghai Jiao Tong University, Shanghai 200240, China} 
	\affiliation{Key Laboratory for Particle Astrophysics and Cosmology (MOE)
		\& Shanghai Key Laboratory for Particle Physics and Cosmology,
		Tsung-Dao Lee Institute  \&	School of Physics and Astronomy, Shanghai Jiao Tong University, Shanghai 200240, China}

	\author { Xin-Yu Du(杜馨雨)}\email{2020dxy@sjtu.edu.cn}

	\affiliation{
		State Key Laboratory of Dark Matter Physics,
		Tsung-Dao Lee Institute \& School of Physics and 
		Astronomy, Shanghai Jiao Tong University, Shanghai 200240, China} 
	\affiliation{Key Laboratory for Particle Astrophysics and Cosmology (MOE)
		\& Shanghai Key Laboratory for Particle Physics and Cosmology,
	Tsung-Dao Lee Institute  \&	School of Physics and Astronomy, Shanghai Jiao Tong University, Shanghai 200240, China}

    \author {Xiao-Gang He(何小刚)}\email{hexg@sjtu.edu.cn}
			
		\affiliation{
		State Key Laboratory of Dark Matter Physics,
		Tsung-Dao Lee Institute \& School of Physics and 
		Astronomy, Shanghai Jiao Tong University, Shanghai 200240, China} 
	\affiliation{Key Laboratory for Particle Astrophysics and Cosmology (MOE)
		\& Shanghai Key Laboratory for Particle Physics and Cosmology,
		Tsung-Dao Lee Institute  \&	School of Physics and Astronomy, Shanghai Jiao Tong University, Shanghai 200240, China}

    \author {Chia-Wei Liu(刘佳韦)}\email{chiaweiliu@ucas.ac.cn}
\affiliation{School of Fundamental Physics and Mathematical Sciences, Hangzhou Institute for Advanced Study, UCAS, Hangzhou 310024, China}
	
	\author {Zi-Yue Zou(邹子月)}\email{ziy\_zou@sjtu.edu.cn}
			
		\affiliation{
		State Key Laboratory of Dark Matter Physics,
		Tsung-Dao Lee Institute \& School of Physics and 
		Astronomy, Shanghai Jiao Tong University, Shanghai 200240, China} 
	\affiliation{Key Laboratory for Particle Astrophysics and Cosmology (MOE)
		\& Shanghai Key Laboratory for Particle Physics and Cosmology,
		Tsung-Dao Lee Institute  \&	School of Physics and Astronomy, Shanghai Jiao Tong University, Shanghai 200240, China}

	\date{\today}

\begin{abstract}
{ Among the charged leptons, the $\tau$ electric dipole moment~($d_\tau$) is the least constrained. We show that the Im[$d_\tau$] imposes strong constraints on new physics that have yet to be discussed. Motivated in particular by the Super Tau-Charm Facility (STCF), which will provide a uniquely clean environment for precision $\tau$-physics, we study the momentum-transfer dependence of $d_\tau(q^2)$ and compare the projected sensitivities of STCF and Belle II. Our analysis shows that an axion-like coupling of the $\tau$ lepton can induce sizable real and imaginary components of the EDM. The predicted EDM values may approach the present experimental sensitivities, making them accessible to future measurements at Belle II and the STCF.}

\end{abstract}

\maketitle

\noindent{{\bf Introduction}}

The electric dipole moment (EDM) of a fundamental particle, $d_f$, violates CP. It is a very powerful probe in studying new physics (NP). Since CP violation beyond the Standard Model (SM) is one of the necessary ingredients to explain why matter dominates over antimatter in our universe, the study of EDMs may therefore shed light on this outstanding problem. Experimental measurements of EDMs have made significant progress since the first measurement of the neutron EDM in 1957~\cite{Smith:1957ht}, though all have yielded null results~\cite{Lamoreaux:1987zz,Muong-2:2008ebm,Abel:2020pzs,Roussy:2022cmp,BESIII:2025vxm}. In the SM, the predicted values of fermion EDMs are many orders of magnitude smaller than the current experimental bounds~\cite{He:1989mbz,Bernreuther:1990jx,Chupp:2017rkp,Yamaguchi:2020eub,Yamaguchi:2020dsy}. Any measurement exceeding the SM prediction would indicate a new source of CP violation, for example, in CP-violating extended Higgs sectors~\cite{Barr:1990vd,Shu:2013uua,Jung:2013hka,Inoue:2014nva}, leptoquark-like scenarios~\cite{He:1992dc,Dorsner:2016wpm,Fuyuto:2018scm,Dekens:2018bci}, supersymmetric models~\cite{Ramsey-Musolf:2006evg,Li:2010ax,Li:2021xmw}.

Due to the short lifetime of the $\tau$ lepton, the measurement of $d_\tau$ is challenging. It has not been investigated to a comparable extent as those of the electron and muon, leaving a significant gap in the lepton sector that requires further studies~\cite{Bernreuther:1996dr,Huang:1996jr,Bernreuther:2021elu,Sun:2024vcd,He:2025ewk,Huang:2025dga,Lu:2025heu,Nakai:2025dmp}.
From a theoretical standpoint, EDMs  require a chirality flip. Since the $\tau$ is the heaviest charged lepton, this chirality-flip scaling naturally makes $d_\tau$ comparatively larger than the corresponding effects for $e$ or $\mu$. Moreover, in a broad class of BSM frameworks, new CP-violating interactions can further enhance $d_\tau$, as CP-violating effects may be larger in the third generation. Consequently, $d_\tau$ may be driven to values large enough to be probed by current or near-future experiments, making improved searches for $d_\tau$ a timely and valuable avenue for discovering or tightly constraining NP.

To probe $d_\tau$, one can generalize it to the case of an off-shell photon, where the generalized $d_\tau(q^2)$ can be studied via the process $\gamma^* (q) \to \tau^+ \tau^-$.
An excellent source of $\gamma^*$ arises from transitions of heavy
quarkonia with the spin-parity $J^{PC} = 1 ^{--}$, such as 
$\psi(3770) \to \gamma^*$ or $\Upsilon \to \gamma^*$~\cite{Bernreuther:1989kc,Bernreuther:1993nd,Belle:2021ybo,Cheng:2025kpp}.
To be explicit, the matrix element $\langle \tau^- | i A_\mu | \tau^- \rangle = e \overline{u}  \Gamma_\mu  u$ can be parameterized in terms of form factors as
\begin{eqnarray}
\begin{split}
    \Gamma^\mu &=F_1(q^2)  \gamma^\mu
+ F_2(q^2) \frac{i \sigma^{\mu\nu} q_\nu}{2 m_\tau}
\\
&+ F_3 (q^2)  \sigma^{\mu\nu} q_\nu \gamma_5  
+ F_4(q^2)  (q^2 \gamma^\mu - q^\mu \slashed{q}) \gamma_5 \,,
\end{split}
\end{eqnarray}
where $d_\tau(q^2) = e F_3(q^2)$, and the ordinary EDM is defined as $d_\tau \equiv d_\tau(0)$.
Notably, it may develop an imaginary part when $q^2 > 0$, a feature that remains largely unexplored and constitutes one of the main focuses of this work. 
The $q^2$ dependence is also physically important, since timelike $d_\tau(q^2)$ can exhibit a nontrivial energy dependence in both its real and imaginary parts. Measuring $d_\tau(q^2)$ at specific timelike momentum transfers therefore provides more information and offers a direct handle on the underlying dynamics.
Belle has reported the most stringent bounds on $d_\tau(q^2)$ at $\sqrt{q^2}=10.58\ \mathrm{GeV}$~\cite{Belle:2021ybo}:  
\begin{eqnarray}
\begin{split}
    \mathrm{Re}[d_{\tau}]   &=  (6.2\pm6.3)\times10^{-18}~e\,\mathrm{cm},\\
\mathrm{Im}[d_{\tau}]    &=  (4.0\pm3.2)\times10^{-18}~e\,\mathrm{cm},
\end{split}
\end{eqnarray}  
which are consistent with the SM prediction of zero. These bounds are expected to be improved by orders of magnitude at future colliders. 
On the other hand, a nonzero $d_\tau$ can induce an electron EDM. Assuming the form factor $F_3(q^2)$ is a real constant, it has been found that $d_\tau(0) \leq 4.1 \times 10^{-19}\,e\,\text{cm}$~\cite{Ema:2022wxd}. However, this result imposes no constraint on $\mathrm{Im}[d_\tau]$. Moreover, it does not constrain NP models whose contributions exhibit a strong $q^2$ dependence at low energies.

Motivated in particular by the Super Tau-Charm Facility (STCF), a major upcoming collider program in China that offers a uniquely clean environment for precision $\tau$ physics in the center-of-mass frame, in which many relevant quantities can be well controlled to get high precision. We analyze the momentum-transfer dependence of the $\tau$ EDM form factor $d_\tau(q^2)$ and compare the projected sensitivities of STCF and Belle II. The expected integrated luminosity for Belle II is about $10~\mathrm{ab}^{-1}$ by around 2034, with the possibility of achieving a larger dataset in the more distant future, enabling a measurement of $d_\tau$~\cite{Belle-II:2018jsg,Belle2LumiProj2024Dec}. 
Since the two experiments operate in different energy regimes, they probe distinct ranges of momentum transfer $q^2$ and can therefore provide complementary information on the $q^2$ dependence of $d_\tau$. 

In this work, from a theoretical perspective, we study whether a sizable $d_\tau(q^2)$, including both its real and imaginary parts, can be generated and tested in the near future. We demonstrate that it can be achieved by a light spin-zero mediator, such as one arising in CP-violating axion-like particle (ALP) phenomenology. For brevity, we will use the term “ALP” to denote this mediator throughout the paper. Future experiments will further probe and constrain this framework.
\\
\\
\noindent{\bf CP violating ALP interaction for $d_\tau(q^2)$}

As a benchmark for our analysis, we  consider a minimal low-energy setup featuring a light spin-zero mediator with CP-violating couplings to $\tau$ leptons. 
We choose it as an illustration because it provides the simplest and most transparent one-loop realization that yields a characteristic timelike threshold behavior, making the physical origin of the $q^2$ dependence explicit. Similar effective couplings are also widely used as standard parametrizations of CP-violating ALP phenomenology in the literature~\cite{DiLuzio:2021jfy,DiLuzio:2023cuk,DiLuzio:2023lmd}. 
In particular, an ALP can be sufficiently light to generate an appreciable $d_\tau$ within this setup.
The same low-energy structure can be matched onto a simple renormalizable UV completion, allowing prospective $d_\tau(q^2)$ measurements to be translated into constraints on underlying parameters. 

The corresponding low-energy effective Lagrangian describing ALP, $a$, couplings to $\tau$ leptons is

\begin{equation}\label{ALP_Lag}
\begin{split}
\mathcal{L} &= \tfrac12 (\partial_\mu a)(\partial^\mu a) - \tfrac12 m_a^2 a^2 \\
& + \bar{\tau}(i\slashed{\partial}-m_\tau)\tau
      + a\,\bar{\tau}(a_a + i b_a \gamma_5)\tau,
\end{split}
\end{equation}
where $a_a,b_a$ are the Yukawa couplings between ALP and $\tau$. Here we use the terminology ALP for our light spin-zero particle. Unlike the usual pseudoscalar ALP, it carries both CP-even and CP-odd couplings, thereby inducing CP violation and generating a nonzero $d_\tau$.

Many studies of ALPs begin with the terms $\tilde{g}_{a\gamma \gamma}\, a F_{\mu\nu}\,\tilde F^{\mu\nu}/4+g_{a\gamma \gamma}\, a F_{\mu\nu}\, F^{\mu\nu}/4$, where $F_{\mu \nu}$ is the electromagnetic field strength tensor\footnote{In most articles concerning ALP, the parameters are written as follows:   $\tilde{g}_{a \gamma \gamma} =c_\gamma \alpha_{\text{em}}/2\pi f_a$ and $ g_{a\gamma \gamma} = \tilde{c}_\gamma \alpha_{\text{em}}/2\pi f_a$. We also note that in much of the ALP literature the notation $g_{a\gamma\gamma}$ often refers to the CP-odd operator $aF\tilde F$ only. In our convention, this corresponds to $\tilde g_{a\gamma\gamma}$.}.
The Yukawa interaction in Eq.~\eqref{ALP_Lag} can likewise reproduce the effects of these two terms via the $\tau$-loop contribution. 
As we have a renormalizable model in mind, which will be discussed later, we will study the effects of these two terms, which are generated by the Yukawa coupling.

\begin{figure*}[ht]
	\centering
	\subfigure[spacelike]{\includegraphics[width=2.8in]{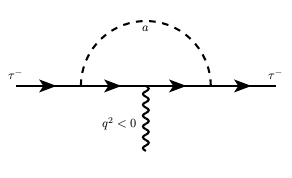}\label{fig:subfig1}}\hspace*{0.2in}
	\subfigure[timelike]{\includegraphics[width=2.8in]{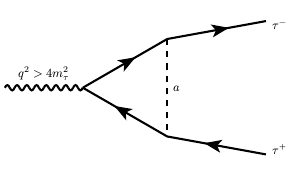}\label{fig:subfig2}}
\caption{One-loop Feynman diagrams contributing to $d_\tau$. The left diagram yields $d_\tau(q^2)$ in the spacelike region ($q^2 < 0$), while the right diagram corresponds to the timelike region ($q^2 > 4m_\tau^2$).}
  \label{fig:1-loop}
\end{figure*}

\begin{figure*}[htbp]
	\centering
	\subfigure[{Re$\left[d_\tau (q^2)\right]$}]{\includegraphics[width=2.8in]{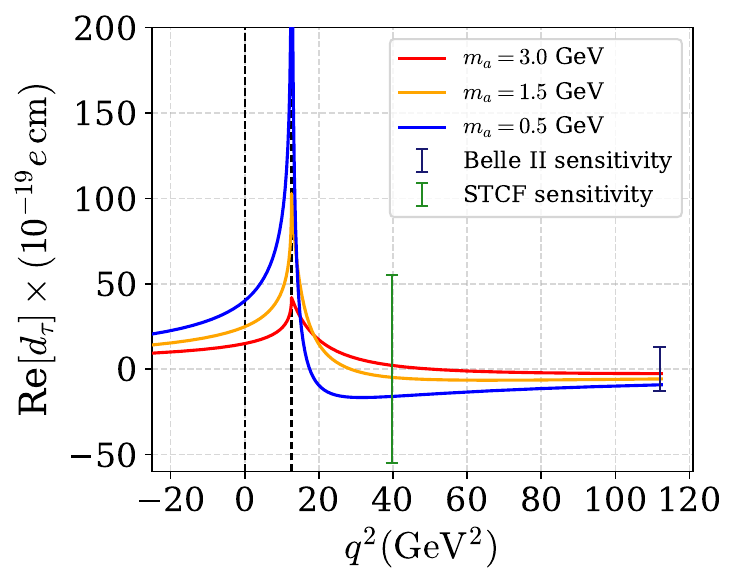}
    }\hspace*{0.2in}
	\subfigure[{Im$\left[d_\tau (q^2)\right]$}]
    {\includegraphics[width=2.8in]{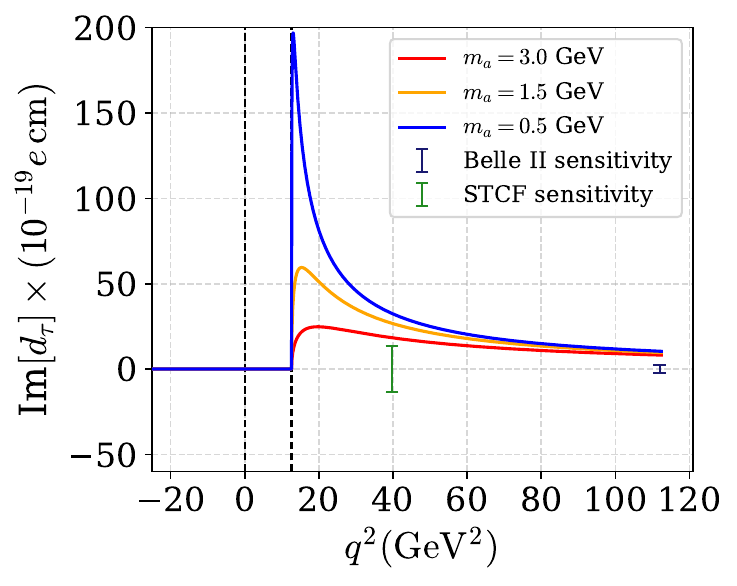}
    }
	\caption{Results for Re$[d_\tau(q^2)]$ and Im$[d_\tau(q^2)]$ in the ALP model are shown for $a_a b_a = 4\times 10^{-2}$, with the EDM scaling proportionally with $a_a b_a$ for other values. Here, the values of $m_a$ are chosen for illustration and are convenient for later discussions. The vertical error bars indicate the projected sensitivity reach at the corresponding timelike energies, $\sqrt{q^2}=10.58~\mathrm{GeV}$ (Belle II) and $\sqrt{q^2}=6.3~\mathrm{GeV}$ (STCF), respectively, at 95\% confidence level (CL). The black dashed lines correspond to $q^2 = 0$ and $q^2 = 4m_\tau^2$.} \label{fig:ALP}
\end{figure*}

Both the real and imaginary parts of $d_\tau(q^2)$ can be generated at the one-loop level, as shown in Fig.~\ref{fig:1-loop}. Evaluating the one-loop diagram, we obtain the EDM form factor, given by  
\begin{widetext}
\begin{eqnarray}\label{oneloopeq}
\begin{split}
    d_\tau (q^2) 
&= 
\frac{e}
{4	\pi ^2 m_\tau (q^2 - 4 m_\tau ^2 ) } 
a_a b_a
\Big [ m_\tau^2
B_0 (  m_\tau ^2, m_\tau ,m_a) - 
  m_\tau ^2 ( B_0 ( q^2 , m_\tau , m_\tau  )  )\\
&+   m _a ^2\left(
  \log (m_\tau / m_a) -   m_\tau ^2 C_0 ( m_\tau ^2, m_\tau ^2,q^2 , m_\tau , m _a , m_\tau  ) 
\right)\Big] \,,
\end{split}
\end{eqnarray}
\end{widetext}
where $B_0$ and $C_0$ are the Passarino-Veltman functions~\cite{Passarino:1978jh, tHooft:1978jhc, Bernreuther:2021elu}. The imaginary part of the EDM can be calculated directly by a simple expression
\begin{eqnarray}
\begin{split}
    \text{Im}[d_{\tau}(q^2)] &=  \frac{e a_a b_a}{4 \pi q^4} \frac{m_\tau m_a^2}{ \beta_\tau^3 } \\
&\times  
\left( \frac{q^2\beta_\tau^2}{m_a^2} -  \log(1 + \frac{q^2 \beta_\tau^2}{m_a^2}) \right) \Theta(q^2-4m_\tau^2),
\end{split}
\label{imdtau}
\end{eqnarray}
where $\beta_\tau  = \sqrt{1 - 4 m_\tau ^2 /q^2 }$. The spacelike region is calculated in Fig.~\ref{fig:subfig1} while the region 
of $q^2\geq 4 m_\tau ^2$ is represented in Fig.~\ref{fig:subfig2}.
The results for $d_\tau(q^2)$ are shown in Fig.~\ref{fig:ALP}.
For completeness of the figure, we extrapolate Re$[d_\tau]$ from both sides into the region $0 < q^2 < 4m_\tau^2$ by continuity, where the $\tau$ pairs are necessarily off-shell. In contrast, the imaginary part is associated with the $\tau\bar{\tau}$ cut; it vanishes in the spacelike channel, and becomes nonzero only above the $\tau^+\tau^-$ production threshold in our analysis.
One can see that $q^2$ indeed affects the value of $d_\tau(q^2)$, implying that specifying the value of $q^2$ is crucial when discussing $d_\tau$. As $q^2$ increases, the corresponding value of EDM decreases in most cases, thereby posing greater challenges for experimental detection. Also, in most cases, a smaller ALP mass corresponds to a larger absolute value for $d_\tau(q^2)$, which implies a greater possibility of being probed experimentally. 

There are many constraints on a light ALP. For $m_a$ below the GeV scale, $\tilde g_{a\gamma\gamma}$ is strongly constrained primarily by cosmological and astrophysical observations~\cite{OHare:2020,ParticleDataGroup:2024cfk}. These limits strongly suppress the allowed couplings, resulting in a correspondingly small EDM. However, a GeV-scale ALP still leaves ample room for generating a sizable $d_\tau(q^2)$, which motivates our choice of $m_a = 0.5,1.5,3$~GeV for illustration in 
Fig.~\ref{fig:ALP}. We restrict to $m_a<2m_\tau$ so that the on-shell decay $a\to\tau^+\tau^-$ is kinematically forbidden. This region is best constrained at BESIII~\cite{BESIII:2022rzz} through the process $\gamma^*  \to a \gamma$
at $q^2 = m_{J/\psi}^2$, given as
\begin{eqnarray}
\begin{aligned}
    	\frac{\mathcal{B}(\gamma^*       \to \gamma a)}{\mathcal{B}(\gamma^*      \to e^+ e^-)} = \frac{q^{ 2}(g_{a\gamma\gamma}^2+\tilde g_{a\gamma\gamma}^2)}{32\pi \alpha_{\text{em}}} \left(1 - \frac{m_a^2}{q^{2}}\right)^3,
\end{aligned}
\end{eqnarray}
and
\begin{widetext}
\begin{eqnarray}\label{gab_relation}
\tilde{g}_{a\gamma\gamma} &=& \frac{b_a m_\tau e^2}{4\pi^2 (q^2 - m_a^2)} \bigg[ P_{\tau}(m_a^2)^2 - P_{\tau}(q^2)^2 \bigg], \\
g_{a\gamma\gamma} &=& \frac{-a_a m_\tau e^2 }{4\pi^2 (q^2 - m_a^2)^2} \Bigg\{ \lambda \Bigg[\left( P_{\tau}(m_a^2) + \frac{2q^2}{\lambda} \sqrt{1 - \frac{4m_\tau^2}{m_a^2}} \right)^2 - \left( P_{\tau}(q^2) + \frac{2q^2\beta_\tau}{\lambda}   \right)^2 \Bigg]  + \frac{16m_\tau^2 q^2 (q^2 + m_a^2)}{m_a^2\lambda} + 4 (m_a^2-q^2)  \Biggr\},\nonumber 
\end{eqnarray}
\end{widetext}
where
$ 
P_{\tau}(q^2) \equiv \log\left(1 + q^2 (\beta_\tau -1 ) /(2m_\tau^2) \right) ,$ and  $
\lambda \equiv q^2 + 4m_\tau^2 - m_a^2. $ 
We also consider the constraints from the OPAL experiment~\cite{OPAL:2002vhf} as recast in Ref.~\cite{Knapen:2016moh} at $q^{2}=0$ from $a\to \gamma \gamma $.

\begin{figure*}[htb]
	\centering
	{\includegraphics[width=4.5in]{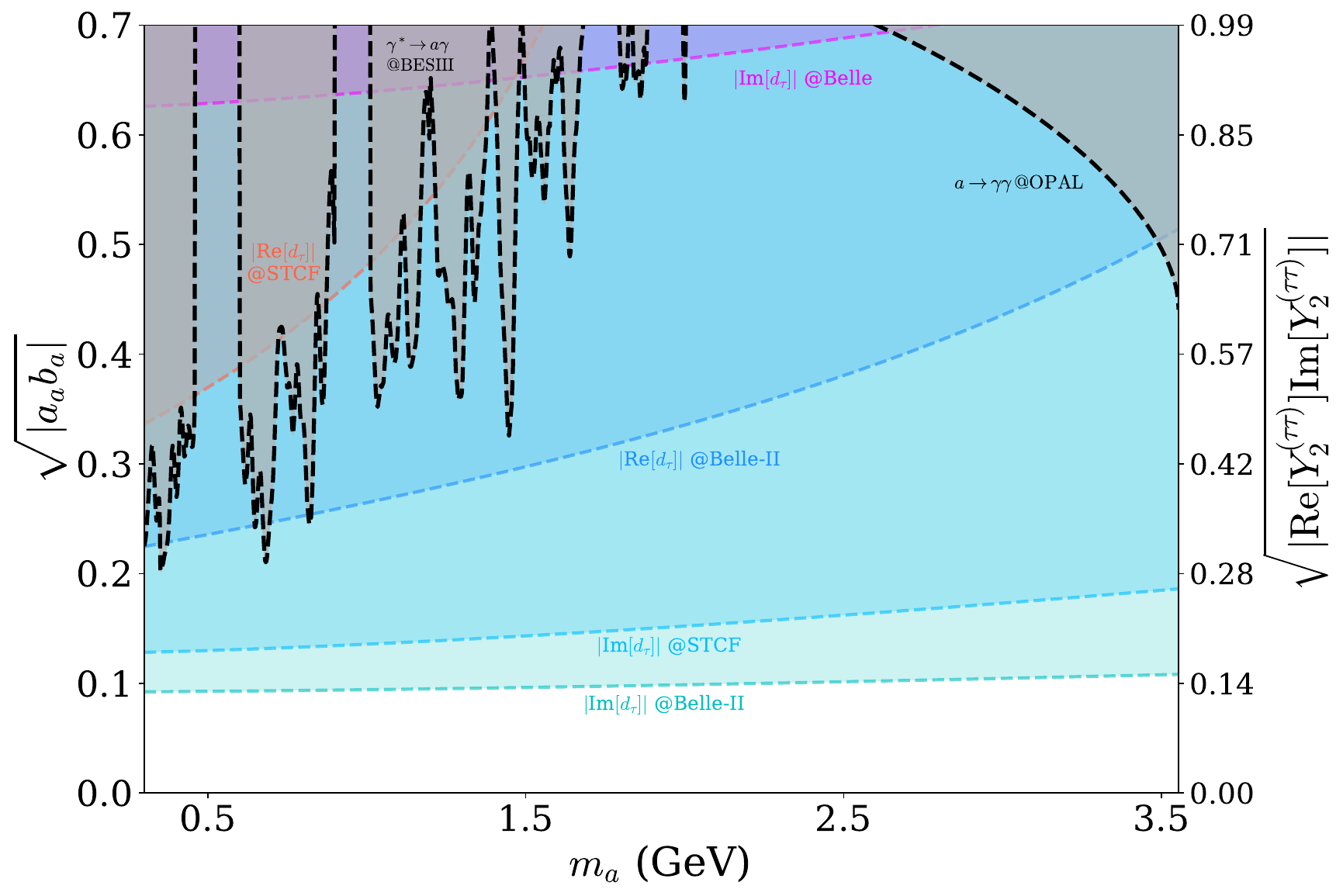}}
	\caption{
    The colored regions represent the parameter space excluded by different experiments. The parameter space above the Belle experimental bound has been ruled out~\cite{Belle:2021ybo}. The curves corresponding to STCF ($\sqrt{q^2}= 6.3~\mathrm{GeV}$ in our benchmarks) and Belle-II ($\sqrt{q^2}= 10.58~\mathrm{GeV}$) indicate the expected bounds from future experimental sensitivities~\cite{He:2025ewk,Lu:2025heu}.
    The gray region 
    is excluded by $\gamma^* \to a \gamma$ limits from BESIII~\cite{BESIII:2022rzz} and the $a\to\gamma\gamma$ limits from the OPAL experiment~\cite{OPAL:2002vhf}, as recast in Ref.~\cite{Knapen:2016moh}. All constraints and projected bounds shown in the figure are consistently presented at 95\%~CL. The right vertical axis shows the corresponding coupling values in the 2HDM realization discussed in Appendix~\ref{app:2HDM}. }
	\label{dtaulimit}
\end{figure*}

The limits from BESIII and OPAL set upper bounds on $\sqrt{ g_{a\gamma\gamma}^2+\tilde g_{a\gamma\gamma}^2}\geq
\sqrt{2 \tilde g_{a\gamma \gamma } g_{a\gamma\gamma}} \propto
\sqrt{a_a b_a}$, where the proportionality coefficients can be directly read from Eq.~\eqref{gab_relation}. We therefore reinterpret these bounds as constraints on $\sqrt{a_a b_a}$ in the CP-violating scenario.
The resulting exclusion is shown as the gray region in Fig.~\ref{dtaulimit}; it is obtained by recasting the BESIII/OPAL limits into our CPV parameterization and is thus directly comparable to the colored regions, which are presented in the same framework.

The expected precision on $d_\tau$ at Belle~II and STCF are collected in Table~\ref{tabledtau}, while the Belle precision is taken from Ref.~\cite{Belle:2021ybo}. For STCF, the quoted precision corresponds to the best achievable sensitivity among the considered center-of-mass energies in Ref.~\cite{He:2025ewk}. These projected sensitivities are also illustrated in Fig.~\ref{fig:ALP} for comparison with our results. The expected experimental constraints on the ALP parameter space are plotted in Fig.~\ref{dtaulimit}. The dashed line represents the experimental limit, while the region above it is excluded. 
From the figure, one can see that Im$[d_\tau]$ imposes a stronger constraint on the parameter space compared to Re$[d_\tau]$.

\begin{table}[h]
    \centering
    \caption{Projected sensitivities to $d_\tau$ at representative timelike energies at 68\% CL~\cite{Belle:2021ybo,He:2025ewk,Lu:2025heu}.}
    \label{tabledtau}
    \begin{tabular}{l c c c}
        \hline\hline
        Collider & $\sqrt{q^2}$ [GeV] &
        $\sigma(\mathrm{Re}\,[d_\tau])$ [$e\,\mathrm{cm}$] &
        $\sigma(\mathrm{Im}\,[d_\tau])$ [$e\,\mathrm{cm}$] \\
        \hline
Belle & 10.58& $6.3 \times 10 ^{-18}$
& $3.2 \times 10 ^{-18}$\\
        {Belle II \footnote{ We adopt the design-target integrated luminosity of 10 $ab^{-1}$ which is projected to be achieved by around 2034.}} & 
        {10.58} &
        { $6.6\times 10^{-19}$} &
        { $1.2\times 10^{-19}$} \\
        STCF & 6.3 &
        $2.8\times 10^{-18}$ &
        $7\times 10^{-19}$ \\
        \hline\hline
    \end{tabular} 
\end{table}

As shown in Table~\ref{tabledtau} and Fig.~\ref{dtaulimit}, Belle II, benefiting from its large integrated luminosity, is expected to provide the most stringent constraints on the imaginary component. However, we stress that the Belle II numbers quoted in Table~\ref{tabledtau} are projections evaluated under the design-target integrated luminosity of 10$~\mathrm{ab}^{-1}$, which is expected to be reached around 2034 as a benchmark value~\cite{Belle-II:2018jsg,Belle2LumiProj2024Dec}. These numbers can be obtained from Ref.~\cite{Lu:2025heu} by rescaling the expected luminosity. At the present stage, the ultimate delivered luminosity remains subject to operational performance and running-time uncertainties.
However, the STCF, operating in the center-of-mass frame, may allow better control of systematic uncertainties with a cleaner event environment and a well-defined initial state. Also, STCF runs at a lower energy (e.g., $\sqrt{q^2}= 6.3~\mathrm{GeV}$ in our benchmarks), probing $d_\tau(q^2)$ at a timelike momentum transfer distinct from Belle II’s $\sqrt{q^2}=10.58~\mathrm{GeV}$. This offers a critical and complementary dataset for constraining the $q^2$ dependence of $d_\tau$.
By measuring $d_\tau(q^2)$ at these distinct energy scales, the combined analysis from Belle II and STCF will be pivotal in establishing the non-trivial $q^2$ dependence of the form factor, which is a hallmark of the loop-induced CP violation predicted in our model.

The ALP interaction in Eq.~(\ref{ALP_Lag}) should be regarded as a minimal showcase of CP-violating dynamics at low energies that generates $d_\tau(q^2)$. However, one may wonder if this can result from a renormalizable, UV-complete model. We indeed find an explicit example based on a CP-violating two-Higgs-doublet model (2HDM). For completeness, some details of this 2HDM realization are collected in Appendix~\ref{app:2HDM}.

Phenomenological studies of ALPs have been extensively explored~\cite{Cao:2010na,Yue:2022ash,Yue:2024xrc,Bao:2025tqs}. We briefly comment on several processes that can be tested experimentally within our model. For example, there are currently no suitable measurements that can be directly recast into meaningful constraints on our parameter space from
$e^+e^- \to \tau^+\tau^- a$, 
$pp \to \tau^+\tau^- a$, or 
$Z \to \tau^+\tau^- \tau^+\tau^-$ . For representative benchmarks, we find that the effects of $a$ are of the order of $\mathrm{BR}(\tau^-\to \mu^- \nu_\tau \bar{\nu}_\mu a)\lesssim4\times10^{-5}$, $\mathrm{BR}(Z\to\tau^+\tau^- a)\lesssim 1 \times 10^{-5}$ and $\mathrm{BR} (Z\to \tau ^+  \tau ^- )_{\mathrm{NP}}/{\rm BR}(Z\to \mu ^+  \mu ^- )\sim 1 \times 10^{-3}$, respectively. They are well below current experimental sensitivities in the parameter space of interest.

Looking ahead, future $Z$ factories such as Future Circular electron-positron Collider (FCC-ee) and Circular Electron Positron Collider (CEPC) are expected to collect a few times of $10^{12}$ $Z$ bosons~\cite{Benedikt:2020ejr,CEPCPhysicsStudyGroup:2022uwl, CEPCStudyGroup:2023quu,Altmann:2025feg,CEPCStudyGroup:2025kmw,Drewes:2025ocf}. 
Our estimates, using the number of $Z$ boson $N_Z \sim 5\times 10^{12} $, indicate that the statistical uncertainty is of order \( 10^{-4} \) for \( \mathrm{BR}(Z \to \tau^+ \tau^- a) \) and \(10^{-3}\) for the ratio \( \mathrm{BR}(Z \to \tau^+ \tau^-)_{\mathrm{NP}}/\mathrm{BR}(Z \to \mu^+ \mu^-) \). 
While this level of precision is insufficient to observe the former, the latter can be experimentally testable and thus provides a viable probe of our model.
\\
\\
\noindent{\bf Conclusion}

In this work, motivated primarily by the STCF’s strong potential for precision $\tau$ physics, we have investigated $d_\tau(q^2)$ as a probe of CP violation beyond the SM, emphasizing the necessity of specifying the experimental energy scale due to its $q^2$ dependence, and have considered NP scenarios. For GeV-scale ALP, one-loop diagrams can generate sizable $\mathrm{Re}[d_\tau(q^2)]$ and $\mathrm{Im}[d_\tau(q^2)]$ within the prospective sensitivities of Belle II and STCF. Furthermore, a CP-violating 2HDM, presented in Appendix~\ref{app:2HDM} as one illustrative example among many possible extensions of the SM, provides a concrete renormalizable framework that can accommodate an ALP and be probed experimentally. By establishing the matching conditions, we showed that future precision measurements of $d_\tau(q^2)$ will critically test the CP-violating sector of extended Higgs models. Current experimental bounds on the effective $a \gamma \gamma$ coupling also exclude substantial regions of the parameter space. Our results underscore the importance of incorporating both real and imaginary components in future experimental analyses and provide benchmarks for upcoming searches at Belle II and STCF, whose measurements are complementary as they correspond to different $q^2$ ranges. Looking further ahead, STCF studies of $d_\tau$ can also serve as a valuable stepping stone towards broader BSM physics opportunities at the long-term high-energy physics facility in China, i.e., the CEPC, as well as at next-generation high-energy colliders internationally, e.g., FCC-ee and LEP3.
\\
\\
\appendix

\section{Two-Higgs-doublet model realization}
\label{app:2HDM}

A renormalizable model of ALP can emerge  from the 2HDM. 
In the Higgs basis, the Yukawa couplings become 	\begin{eqnarray}
\begin{aligned}
    	-\mathcal{L}_Y=&\bar{L_L}Y_1\Phi_1l_R+\bar{L_L}Y_2\Phi_2l_R+h.c.,
\end{aligned}
	\label{yukawa}
\end{eqnarray}
and the scalar potential of the two Higgs doublets $\Phi_{1,2}$ is~\cite{Wu:1994ja,Gunion:2002zf}
\begin{widetext}
\begin{eqnarray}
\begin{aligned}
\hspace{-1em}
V(\Phi_1,\Phi_2)
&= m_{11}^2\Phi_1^\dagger\Phi_1+m_{22}^2\Phi_2^\dagger\Phi_2 - (m_{12}^2 \Phi_1^\dagger \Phi_2 + \text{h.c.})  +\frac{1}{2}\lambda_1\left(\Phi_1^\dagger\Phi_1\right)^2+\frac{1}{2}\lambda_2\left(\Phi_2^\dagger\Phi_2\right)^2+\lambda_3\left(\Phi_1^\dagger\Phi_1\right)\left(\Phi_2^\dagger\Phi_2\right)\\
&+\lambda_4\left(\Phi_1^\dagger\Phi_2\right)\left(\Phi_2^\dagger\Phi_1\right) +\left(\frac{1}{2}\lambda_5\left(\Phi_1^\dagger\Phi_2\right)^2+\left(\lambda_6\Phi_1^\dagger\Phi_1+\lambda_7\Phi_2^\dagger\Phi_2\right)\left(\Phi_1^\dagger\Phi_2\right)+\mathrm{c.c.}\right),
\end{aligned}
\end{eqnarray}
\end{widetext}
where
\begin{eqnarray}
\begin{aligned}
\hspace{-1.5em}
    \Phi_1=&\begin{pmatrix}G^+\\\frac{1}{\sqrt{2}}\left(v+h +iG^0\right)\end{pmatrix}, \Phi_2=\!\begin{pmatrix}H^+\\\frac{1}{\sqrt{2}}\left(-H+iA \right)\end{pmatrix}\!.
\end{aligned}
\end{eqnarray} 
$H$ and $A$ are the new neutral Higgs particles. Here one can diagonalize $Y_{1 }$ to obtain the mass matrix, and the rotation matrix is absorbed by a redefinition of the lepton fields. A rephrasing of $\Phi_2$ would in general reshuffle phases between Yukawa coupling and potential parameters, but here we fix the basis in which the potential is CP conserving and keep the physical CP-violating phase in the Yukawa coupling sector. We work in the limit that all $\lambda_i$ are real for clarity of the analysis. The mass matrix can be written as
\begin{widetext}
\begin{eqnarray}
	\frac{M^2}{v^2}=\begin{pmatrix}\lambda_1&\mathrm{Re}[\lambda_6]&-\mathrm{Im}[\lambda_6]\\\\\mathrm{Re}[\lambda_6]&m_{22}^2+\frac{1}{2}(\lambda_3+\lambda_4+\mathrm{Re}[\lambda_5])v^2&-\frac{1}{2}\mathrm{Im}[\lambda_5]\\\\-\mathrm{Im}[\lambda_6]&-\frac{1}{2}\mathrm{Im}[\lambda_5]&m_{22}^2+\frac{1}{2}(\lambda_3+\lambda_4-\mathrm{Re}[\lambda_5])v^2\end{pmatrix}.
\end{eqnarray}
\end{widetext}
In the alignment limit, $\lambda_6=0$, such that there is no mixing between $h$ and the other neutral Higgs bosons; accordingly, $h$ is identified with the SM Higgs boson. Note that $d_\tau(q^2)$ originates from the leptonic Yukawa coupling. Therefore, the alignment limit is consistent with nonzero CP violation in the Yukawa sector. Also, since the scalar potential is CP-conserving in our setup (with $\lambda_5$ real), $h$ contains no CP-odd admixture, and the stringent LHC limits on the SM Higgs are automatically satisfied at tree-level.

Without loss of generality, we assume that the contribution to $d_\tau$ is dominated by $A$. Similar contributions from $H$ exist, which are proportional to $m_A^2 / m_H^2$ and thus negligible. Keeping only $Y_2^{\tau\tau}$ non-zero, $A$ then plays the role of the ALP, $a$, in Eq.~\ref{ALP_Lag} with the identification
	\begin{eqnarray}
		\begin{aligned}
			a_a&=-\operatorname{Im}[Y_2^{\tau \tau }]/\sqrt{2}, ~~~ 
			b_a =\operatorname{Re}[Y_2^{\tau \tau }]/\sqrt{2}.
		\end{aligned}
\end{eqnarray} 
Therefore, the contributions of $A$ to $d_\tau(q^2)$ can be obtained from the results presented in the main text. The constrained parameter space is identical to that of the general ALP scenario shown in Fig.~\ref{dtaulimit}. For the 2HDM case, the corresponding couplings are indicated on the right vertical axis. The existing constraints are already taken into account~\cite{LEPHiggsWorkingGroupforHiggsbosonsearches:2001ogs,Gunion:2002zf,Haber:2010bw,ALEPH:2013htx,ATLAS:2023tkt}, however, they do not affect our results here.

If $m_A$ is around the electroweak scale, the resulting $d_\tau$ is suppressed by several orders of magnitude, and remains well below the reach of near-future experiments~\cite{Bernreuther:2021elu}. 
Hence, an observable effect can only be seen with the mass of $A$ to be around GeV level. 

A detailed analysis of the viable parameter space of the 2HDM, including constraints from the Higgs potential, will be presented elsewhere.

\begin{acknowledgments}
This work is supported in part by the National Key Research and Development Program of China under Grant
No. 2020YFC2201501, by the Fundamental Research
Funds for the Central Universities, by National Natural Science Foundation of P.R. China (No.12090064,
12205063, 12375088 and W2441004).
\end{acknowledgments}

\end{CJK*}

\begin{thebibliography}{99}\footnotesize
\itemsep=-1pt plus.2pt minus.2pt

\bibitem{Smith:1957ht}
J.~H.~Smith, E.~M.~Purcell and N.~F.~Ramsey 1957
{\it Phys. Rev.} \textbf{108} 120-122 

\bibitem{Lamoreaux:1987zz}
S.~K.~Lamoreaux, J.~P.~Jacobs, B.~R.~Heckel, F.~J.~Raab and N.~Fortson 1987
{\it Phys. Rev. Lett.} \textbf{59} 2275-2278

\bibitem{Muong-2:2008ebm}
G.~W.~Bennett \textit{et al.} [Muon (g-2)] 2009
{\it Phys. Rev. D} \textbf{80} 052008 

\bibitem{Abel:2020pzs}
C.~Abel, S.~Afach, N.~J.~Ayres, C.~A.~Baker, G.~Ban, G.~Bison, K.~Bodek, V.~Bondar, M.~Burghoff and E.~Chanel \textit{et al.}
2020
{\it Phys. Rev. Lett.} \textbf{124} 081803 

\bibitem{Roussy:2022cmp}
T.~S.~Roussy, L.~Caldwell, T.~Wright, W.~B.~Cairncross, Y.~Shagam, K.~B.~Ng, N.~Schlossberger, S.~Y.~Park, A.~Wang and J.~Ye \textit{et al.}
2023 {\it Science} \textbf{381} adg4084 

\bibitem{BESIII:2025vxm}
M.~Ablikim \textit{et al.} [BESIII]
2025 arXiv:2506.19180 [hep-ex]

\bibitem{He:1989mbz}
X.~G.~He, B.~H.~J.~McKellar and S.~Pakvasa,
1989
{\it Int. J. Mod. Phys. A} \textbf{4}, 5011 
; Erratum: [1991 {\it Int. J. Mod. Phys. A} \textbf{6}, 1063-1066]

\bibitem{Bernreuther:1990jx}
W.~Bernreuther and M.~Suzuki
1991 {\it Rev. Mod. Phys.} \textbf{63}, 313-340 
; Erratum: [1992 {\it Rev. Mod. Phys.} \textbf{64} 633 ]

\bibitem{Chupp:2017rkp}
T.~Chupp, P.~Fierlinger, M.~Ramsey-Musolf and J.~Singh
2019
{\it Rev. Mod. Phys.} \textbf{91} 015001 

\bibitem{Yamaguchi:2020eub}
Y.~Yamaguchi and N.~Yamanaka
2020 {\it Phys. Rev. Lett. }\textbf{125} 241802 

\bibitem{Yamaguchi:2020dsy}
Y.~Yamaguchi and N.~Yamanaka
2021 {\it Phys. Rev. D }\textbf{103} 013001 

\bibitem{Barr:1990vd}
S.~M.~Barr and A.~Zee
1990 {\it Phys. Rev. Lett. }\textbf{65} 21-24 
Erratum: [1990 {\it Phys. Rev. Lett.} \textbf{65} 2920 ]

\bibitem{Shu:2013uua}
J.~Shu and Y.~Zhang
2013 {\it Phys. Rev. Lett. }\textbf{111} 091801 

\bibitem{Jung:2013hka}
M.~Jung and A.~Pich
2014 {\it JHEP} \textbf{04} 076 

\bibitem{Inoue:2014nva}
S.~Inoue, M.~J.~Ramsey-Musolf and Y.~Zhang
2014 {\it Phys. Rev. D} \textbf{89} 115023 

\bibitem{He:1992dc}
X.~G.~He, B.~H.~J.~McKellar and S.~Pakvasa
1992 {\it Phys. Lett. B} \textbf{283} 348-352

\bibitem{Dorsner:2016wpm}
I.~Dor{\v{s}}ner, S.~Fajfer, A.~Greljo, J.~F.~Kamenik and N.~Ko{\v{s}}nik
2016 {\it Phys. Rept.} \textbf{641} 1-68 

\bibitem{Fuyuto:2018scm}
K.~Fuyuto, M.~Ramsey-Musolf and T.~Shen
2019 {\it Phys. Lett. B} \textbf{788} 52-57 

\bibitem{Dekens:2018bci}
W.~Dekens, J.~de Vries, M.~Jung and K.~K.~Vos
2019 
{\it JHEP} \textbf{01} 069 

\bibitem{Ramsey-Musolf:2006evg}
M.~J.~Ramsey-Musolf and S.~Su
2008 
{\it Phys. Rept.} \textbf{456} 1-88 

\bibitem{Li:2010ax}
Y.~Li, S.~Profumo and M.~Ramsey-Musolf
2010 {\it JHEP} \textbf{08} 062 

\bibitem{Li:2021xmw}
S.~Li, Y.~Xiao and J.~M.~Yang
2022 {\it Nucl. Phys. B} \textbf{974} 115629 

\bibitem{Bernreuther:1996dr}
W.~Bernreuther, A.~Brandenburg and P.~Overmann
1997 {\it Phys. Lett. B} \textbf{391} 413-419;
 Erratum: [1997 {\it Phys. Lett. B} \textbf{412} 425-425 ]

\bibitem{Huang:1996jr}
T.~Huang, W.~Lu and Z.~j.~Tao
1997 {\it Phys. Rev. D} \textbf{55} 1643-1652 

\bibitem{Bernreuther:2021elu}
W.~Bernreuther, L.~Chen and O.~Nachtmann
2021 {\it Phys. Rev. D} \textbf{103} 096011 

\bibitem{Sun:2024vcd}
X.~Sun, Y.~Wu and X.~Zhou
2025 {\it Chin. Phys.} \textbf{49} 113001 

\bibitem{He:2025ewk}
X.~G.~He, C.~W.~Liu, J.~P.~Ma, C.~Yang and Z.~Y.~Zou
2025 {\it JHEP} \textbf{04} 001 

\bibitem{Huang:2025dga}
Z.~L.~Huang and X.~G.~He
2025 {\it JHEP} \textbf{07} 205 

\bibitem{Lu:2025heu}
P.~C.~Lu, Z.~G.~Si and H.~Zhang
2025 {\it Phys. Rev. D} \textbf{112} 075039 

\bibitem{Nakai:2025dmp}
Y.~Nakai, Y.~Shigekami, P.~Sun and Z.~Zhang
2025 arXiv:2508.05935 [hep-ph]

\bibitem{Bernreuther:1989kc}
W.~Bernreuther and O.~Nachtmann
1989 {\it Phys. Rev. Lett. }\textbf{63} 2787; Erratum: [1990 {\it Phys. Rev. Lett.} \textbf{64} 1072]

\bibitem{Bernreuther:1993nd}
W.~Bernreuther, O.~Nachtmann and P.~Overmann,
1993 {\it Phys. Rev. D} \textbf{48} 78-88 

\bibitem{Belle:2021ybo}
K.~Inami \textit{et al.} [Belle]
2022 {\it JHEP} \textbf{04} 110 

\bibitem{Cheng:2025kpp}
H.~Y.~Cheng, Z.~H.~Guo, X.~G.~He, Y.~Hou, X.~W.~Kang, A.~Kupsc, Y.~Y.~Li, L.~Liu, X.~R.~Lyu and J.~P.~Ma \textit{et al.}
2025 arXiv:2502.08907 [hep-ex]

\bibitem{Ema:2022wxd}
Y.~Ema, T.~Gao and M.~Pospelov
2022 {\it Phys. Lett. B} \textbf{835} 137496 

\bibitem{Belle-II:2018jsg}
E.~Kou \textit{et al.} [Belle-II]
2019 {\it PTEP} \textbf{2019} 123C01; Erratum: [2020 {\it PTEP} \textbf{2020} 029201]

\bibitem{Belle2LumiProj2024Dec}
SuperKEKB/Belle~II,
``Luminosity Projection (Dec 2024),''
Online PDF, available at \url{https://public.belle2.org/downloads/LuminosityProjection_2024Dec.pdf}
(accessed 24 Jan 2026).

\bibitem{DiLuzio:2021jfy}
L.~Di Luzio
2022 {\it PoS} \textbf{EPS-HEP2021} 513 

\bibitem{DiLuzio:2023cuk}
L.~Di Luzio, G.~Levati and P.~Paradisi
2024 {\it JHEP} \textbf{02} 020

\bibitem{DiLuzio:2023lmd}
L.~Di Luzio, H.~Gisbert, G.~Levati, P.~Paradisi and P.~S{\o}rensen
2023 arXiv:2312.17310 [hep-ph]

\bibitem{Passarino:1978jh}
G.~Passarino and M.~J.~G.~Veltman,
1979 {\it Nucl. Phys. B} \textbf{160} 151-207

\bibitem{tHooft:1978jhc}
G.~'t Hooft and M.~J.~G.~Veltman
1979 {\it Nucl. Phys. B} \textbf{153} 365-401 

\bibitem{OHare:2020}
C.~O'Hare,
Zenodo 2020
\url{https://cajohare.github.io/AxionLimits/}.

\bibitem{ParticleDataGroup:2024cfk}
S.~Navas \textit{et al.} [Particle Data Group] 2024
{\it Phys. Rev. D} \textbf{110} 030001

\bibitem{BESIII:2022rzz}
M.~Ablikim \textit{et al.} [BESIII]
2023 {\it Phys. Lett. B} \textbf{838} 137698

\bibitem{OPAL:2002vhf}
G.~Abbiendi \textit{et al.} [OPAL]
2003 {\it Eur. Phys. J. C} \textbf{26} 331-344 

\bibitem{Knapen:2016moh}
S.~Knapen, T.~Lin, H.~K.~Lou and T.~Melia
2017 {\it Phys. Rev. Lett.} \textbf{118} 171801 

\bibitem{Cao:2010na}
J.~Cao, Z.~Heng and J.~M.~Yang
2010 {\it JHEP} \textbf{11} 110

\bibitem{Yue:2022ash}
C.~X.~Yue, S.~Yang, H.~Wang and N.~Zhang
2022 {\it Phys. Rev. D} \textbf{105} 115027

\bibitem{Yue:2024xrc}
C.~X.~Yue, X.~Y.~Li and X.~C.~Sun
2024 {\it Eur. Phys. J. C} \textbf{84} 1033

\bibitem{Bao:2025tqs}
S.~s.~Bao, Y.~Ma, Y.~Wu, K.~Xie and H.~Zhang
2025 {\it JHEP} \textbf{10} 122

\bibitem{Benedikt:2020ejr}
M.~Benedikt, A.~Blondel, P.~Janot, M.~Mangano and F.~Zimmermann
2020 {\it Nature Phys.} \textbf{16} 402-407

\bibitem{CEPCPhysicsStudyGroup:2022uwl}
H.~Cheng \textit{et al.} [CEPC Physics Study Group]
2022 arXiv:2205.08553 [hep-ph]

\bibitem{CEPCStudyGroup:2023quu}
W.~Abdallah \textit{et al.} [CEPC Study Group]
2024 {\it Radiat. Detect. Technol. Methods} \textbf{8} 1-1105; Erratum: [2025 {\it Radiat. Detect. Technol. Methods} \textbf{9} 184-192]

\bibitem{Altmann:2025feg}
J.~Altmann, P.~Skands, A.~Desai, W.~Mitaroff, S.~Pl{\"a}tzer, D.~Dobur, K.~Skovpen, M.~Drewes, G.~Durieux and Y.~Georis \textit{et al.}
2025 arXiv:2506.15390 [hep-ex]

\bibitem{CEPCStudyGroup:2025kmw}
S.~P.~Adhya \textit{et al.} [CEPC Study Group]
2025 arXiv:2510.05260 [hep-ex]

\bibitem{Drewes:2025ocf}
M.~Drewes, J.~Klari{\'c} and Y.~Z.~Li
2025 arXiv:2511.23461 [hep-ph]

\bibitem{Wu:1994ja}
Y.~L.~Wu and L.~Wolfenstein
1994 {\it Phys. Rev. Lett.} \textbf{73} 1762-1764

\bibitem{Gunion:2002zf}
J.~F.~Gunion and H.~E.~Haber
2003 {\it Phys. Rev. D} \textbf{67} 075019

\bibitem{LEPHiggsWorkingGroupforHiggsbosonsearches:2001ogs}
 [LEP Higgs Working Group for Higgs boson searches, ALEPH, DELPHI, L3 and OPAL]
2001 arXiv:hep-ex/0107031 [hep-ex]

\bibitem{Haber:2010bw}
H.~E.~Haber and D.~O'Neil
2011 {\it Phys. Rev. D} \textbf{83} 055017

\bibitem{ALEPH:2013htx}
G.~Abbiendi \textit{et al.} [ALEPH, DELPHI, L3, OPAL and LEP]
2013 {\it Eur. Phys. J. C} \textbf{73} 2463

\bibitem{ATLAS:2023tkt}
G.~Aad \textit{et al.} [ATLAS]
2023 {\it Phys. Lett. B} \textbf{842} 137963 

\end{thebibliography}
\end{document}